\documentclass[english]{IEEEtran}
\usepackage[T1]{fontenc}
\usepackage[latin9]{inputenc}
\usepackage{float}
\usepackage{url}
\usepackage{graphicx}

\makeatletter

\providecommand{\tabularnewline}{\\}

\@ifundefined{showcaptionsetup}{}{%
 \PassOptionsToPackage{caption=false}{subfig}}
\usepackage{subfig}
\makeatother

\usepackage{babel}
\usepackage{listings}
\usepackage{listings-golang}
\usepackage{color}

\definecolor{arsenic}{rgb}{0.23, 0.27, 0.29}

\begin{document}

\title{SCTP in Go}

\author{Olivier Van Acker\\
Department of Computer Science and Information Systems\\
 Birkbeck University of London\\
 London, United Kingdom\\
 Email: ovanac01@mail.bbk.ac.uk}
\maketitle
\begin{abstract}
This paper describes a successful attempt to combine
two relatively new technologies: Stream Control Transmission
Protocol (SCTP) and the programming language Go,
achieved by extending the existing Go network library with
SCTP.
SCTP is a reliable, message-oriented transport layer protocol,
similar to TCP and UDP. It offers sequenced delivery of messages
over multiple streams, network fault tolerance via multihoming
support, resistance against flooding and masquerade attacks
and congestion avoidance procedures. It has improvements over
wider-established network technologies and is gradually gaining
traction in the telecom and Internet industries.
Go is an open source, concurrent, statically typed, compiled
and garbage-collected language, developed by Google Inc. Go's
main design goals are simplicity and ease of use and it has a
syntax broadly similar to C. Go has good support for networked
and multicore computing and as a system language is often used
for networked applications, however it doesn't yet support SCTP.
By combining SCTP and Go, software engineers can exploit the
advantages of both technologies. The implementation of SCTP
extending the Go network library was done on FreeBSD and
Mac OS X - the two operating systems that contain the most up
to date implementation of the SCTP specification.
\end{abstract}


\section{Introduction}

This paper will look into combining two relatively new technologies:
a network protocol called SCTP and the programming language Go. Both
technologies claim to offer improvements over existing technologies:
SCTP does away with the limitations of the widely used TCP protocol
\cite{rfc4960}; and Go was designed with simplicity and minimized
programmer effort in mind, thereby preventing a forest of features
getting in the way of program design: Less is exponentially more \cite{pike_command_2012}.
The current version of the Go network library does not support the
SCTP protocol and this paper examines how easy it is to extend the
Go network library with this protocol. The work in this paper is based
on the dissertation submitted for an MSc in Computer Science at
Birkbeck University in London and is available as an open source project.

\subsection{Relevance}

After ten years SCTP as a technology is becoming more and more relevant.
One notable implementation is the use of SCTP as a data channel in
the Web Real-Time Communication (WebRTC) standard \cite{rtcweb_sctp},
a HTML 5 extension to enable real time video and audio communication
in browsers. Google inc. and the Mozilla Foundation are each planning
to release a browser this year implementing the WebRTC standard. Go
as a language is very new and it is too early to say what impact it
will have. It does however receive a lot of media attention since
it is developed by Google. It is also growing in popularity because
of its ease of use as a concurrent system language \cite{oreilly_why_go}.

\subsection{Outline }
Section II presents an overview of Go and SCTP, followed by (section III) a description of how the TCP socket API is integrated in the Go networking library. This is a starting point for the design of an SCTP extension to the network library in Go, described in section IV. Section V explains the implementation. Section VI analysis the results and VII concludes.

\section{Technology overview}

In this section I will give an overview of the main features of SCTP
and Go.

\subsection{SCTP }

SCTP was initially developed in response to the demands of the telecoms
industry, which was not satisfied with the reliability and performance
of TCP \cite[p. 15]{stewart_stream_2001}. During the design phase
the decision was made to make SCTP a less telephony-centric IP protocol
\cite[p. 16]{stewart_stream_2001} so that it could also be used for
more generic data transport purposes. 

\subsubsection{Comparison with TCP}

It is fruitful to compare TCP and SCTP, as TCP is the most widely-used
protocol \cite{sctp_userspace} and SCTP is very similar to it:

\subsubsection{Multihoming}

A host is said to be multihomed if it has multiple IP addresses which
can be associated with one or more physical interfaces connected to
the same or different networks\cite{rfc1122}. TCP can only bind to
one network address at each end of the connection. In the event of
a network failure there is no way to stay connected and send the data
over another physical path without setting up a new connection. SCTP
natively supports multihoming at the transport layer. This makes it
possible to connect and transfer data to and from multihomed hosts,
so that when one network path fails, the connection seamlessly fails
over using the remaining paths. And together with concurrent multipath
transfer (CMT) \cite{cmt_sctp} it is possible to increase data throughput
by transferring data simultaneously over multiple links. 

\subsubsection{In-built message boundaries}

In TCP there are no markers to indicate the start or end of a specific
piece of data (a user message). All data sent over the socket is converted
into a byte stream and an application must add its own mechanism to
be able to reconstruct the messages. In SCTP the message boundaries
of the data sent are preserved in the protocol. In the event of the
message being larger than the maximum package size a notofication
is sent to the application layer more is on its way.

\subsubsection{Protection against Denial of Service (DOS) attacks}

Another disadvantage of TCP is that it is open to \textquoteright SYN
flood attack\textquoteright , a specific type of attack which can
drain the server of resources, resulting in a denial of service (nothing
else can connect). To prevent this, SCTP uses a four-way handshake
to establish the connection, whereas TCP only uses a three-way handshake.
With a four-way handshake the originator has to double-acknowledge
itself (\textquotedblright is it really you?\textquotedblright ) by
resending a cookie it previously received from the destination server
before that server will assign resources to the connection. This prevents
timeouts on the server side and thus makes this type of denial of
service impossible. To reduce start up delay, actual data can also
be sent during the second part of the handshake.

\begin{figure}[H]
\caption{Initiating a network connection}
\centering{}\subfloat[TCP handshake]{\includegraphics[scale=0.3]{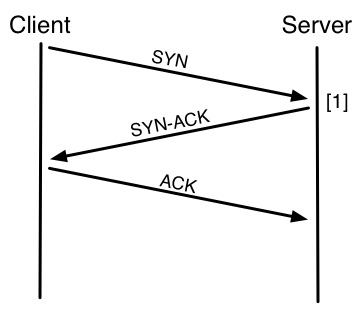}

}\ \subfloat[SCTP handshake]{\includegraphics[scale=0.3]{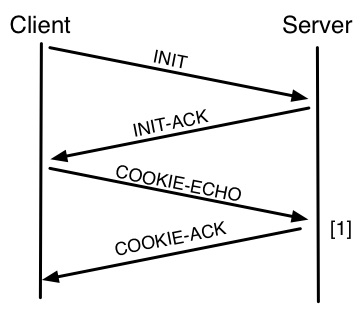}

}
\end{figure}

\subsubsection{SCTP Multistreaming}

\begin{figure}

\caption{Head of line blocking\label{fig:Head-of-line}}
\noindent\begin{minipage}[t]{1\columnwidth}%
\begin{center}
\includegraphics[scale=0.45]{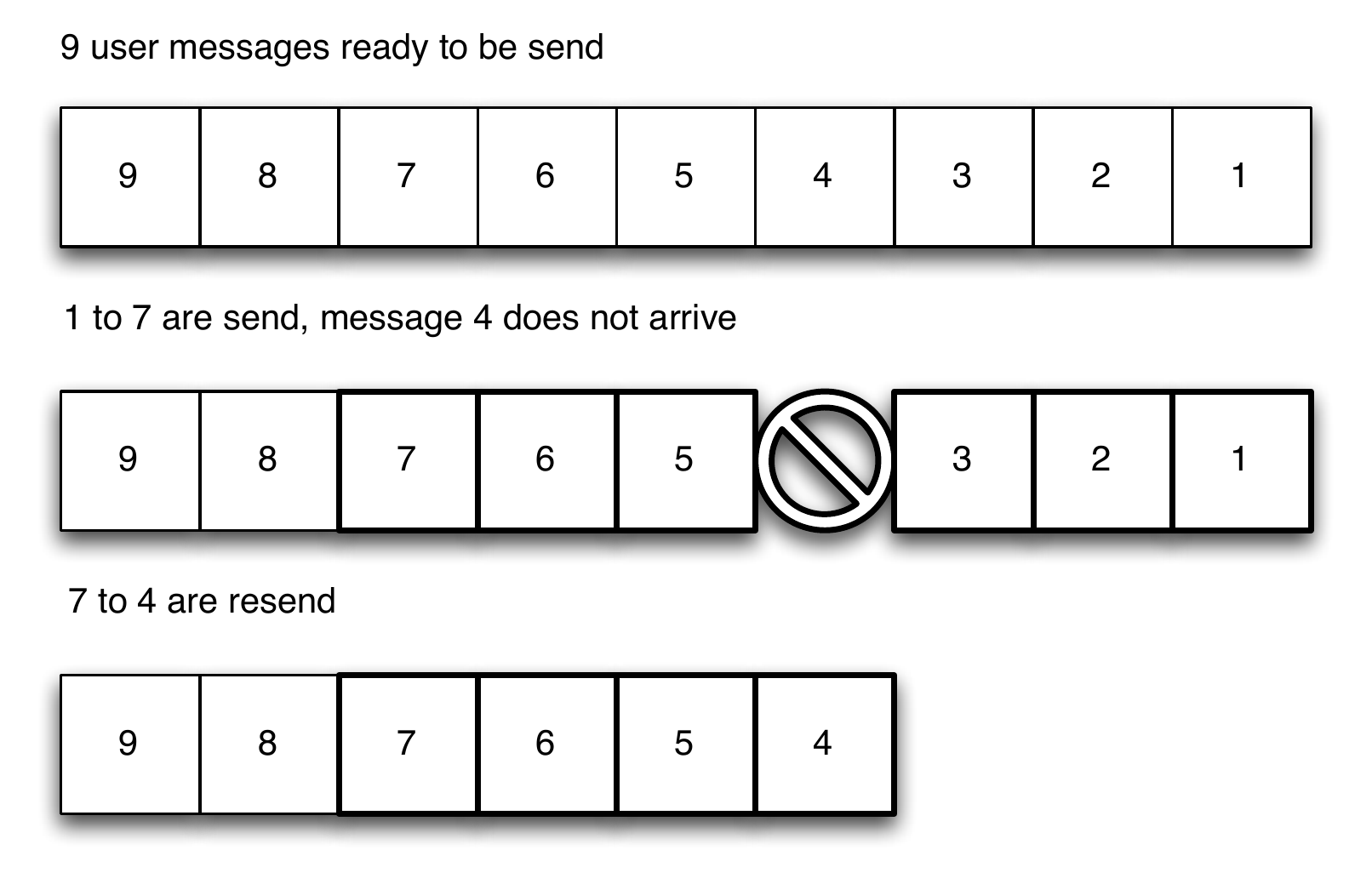}
\par\end{center}%
\end{minipage}

\end{figure}

SCTP user messages in a single SCTP socket connection can be sent
and delivered to the application layer over independent streams. In
case of two sets of user messages A and B, each set delivered sequentially,
the messages of set A can be sent over a different stream than B.
And in case a messages in set A gets missing and part of the sequence
needs to be resent this will not effect the data of set B if it is
sent over a different stream. This tackels the 'head of line blocking'
problem (figure \ref{fig:Head-of-line}) where messages already delivered
need to be redelivered because they have to arrive in order and becuase
one message did not arrive. 

\subsubsection{Associations}

\begin{figure}[h]

\caption{Associations and streams\label{fig:Associations-and-streams}}

\noindent\begin{minipage}[t]{1\columnwidth}%
\begin{center}
\includegraphics[scale=0.35]{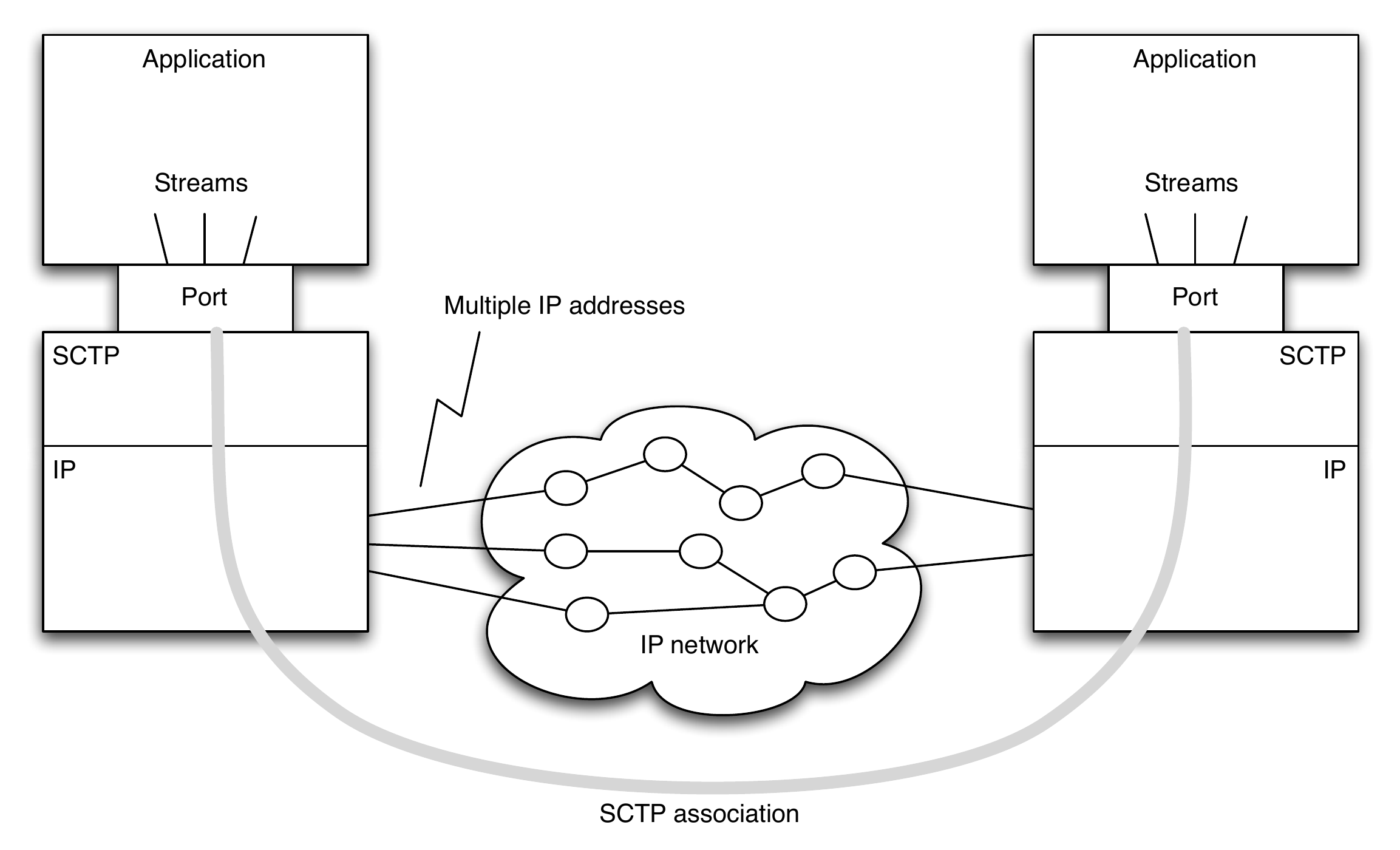}
\par\end{center}%
\end{minipage}

\end{figure}

A connection between a SCTP server and client is called an association.
An association gets initiated by a request from a SCTP client. So
a listening server can accept incoming requests from multiple clients.
Messages sent over the assocation have an association id attached
to them. to make it possible to know where they come from. Within
a single association you can have multiple streams for data transmission
(See figure \ref{fig:Associations-and-streams})

\subsubsection{Two programming models\label{subsec:Two-programming-models}}

SCTP has two interfaces for implementation in a networked application:
one-to-one and one-to-many. A single socket in a one-to-many model
can have multiple incoming associations, meaning that multiple clients
can connect to a single server listening on a single socket. The one-to-one
model can only have a single association per socket. The one-to-one
model is for easy migration of existing TCP applications, it maps
one-to-one to the system calls TCP makes to establish a connection.
But it can only have one connection per association, and thus only
a single client can connect to a server. The one-to-one interface
makes migrating an existing application a relatively painless exercise.
If you want to upgrade your existing TCP application to a one-to-many
SCTP application, significant retooling is needed. \cite[p. 267]{stevens_unix_2003}

\begin{figure}[h]
\caption{Socket API\label{fig:Socket-API}}

\noindent\begin{minipage}[t]{1\columnwidth}%
\begin{center}
\includegraphics[scale=0.4]{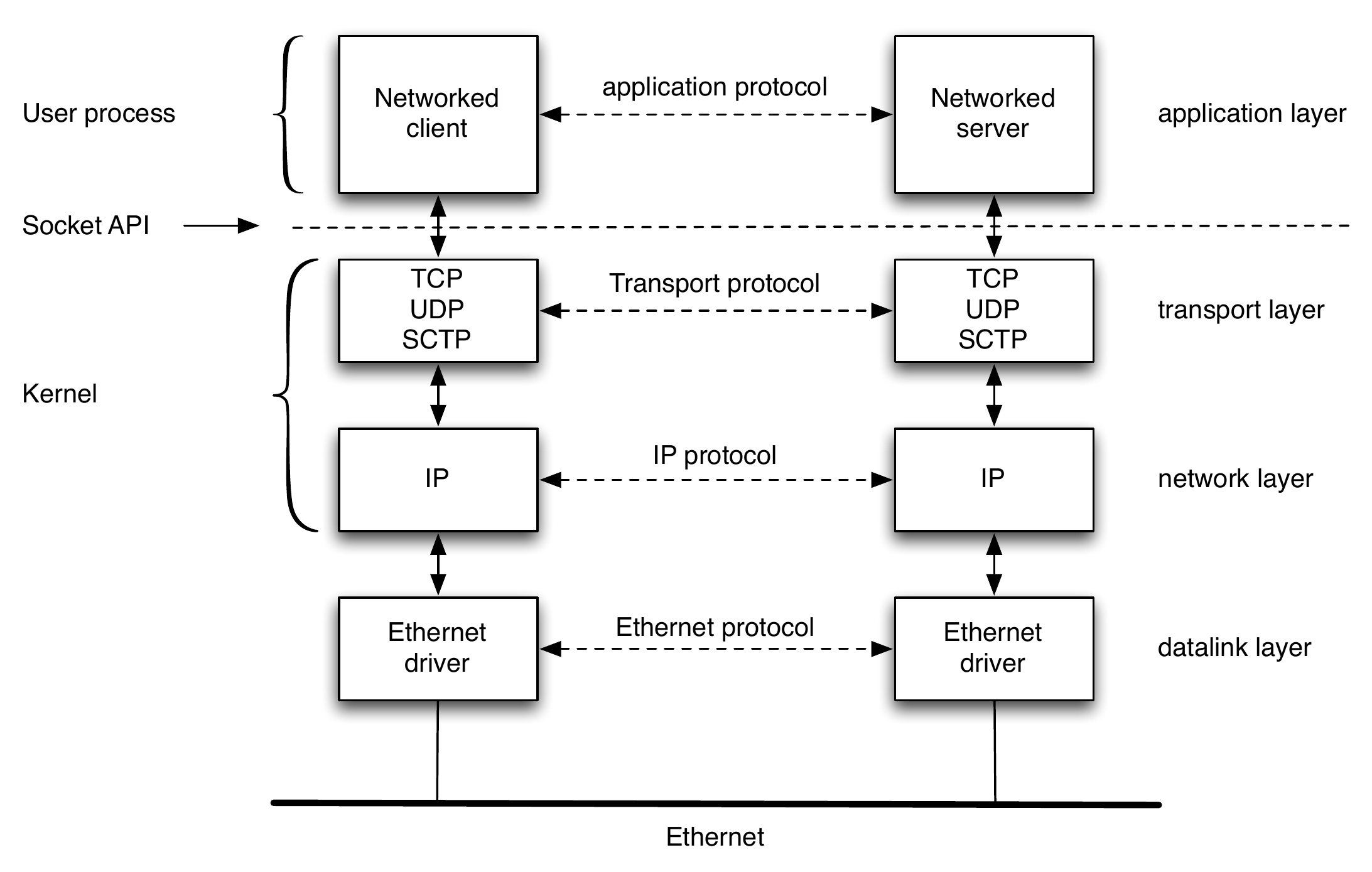}
\par\end{center}%
\end{minipage}
\end{figure}

\subsubsection{Socket API}

Figure \ref{fig:Socket-API} is an overview of two applications communicating
over an IP network using a transport protocol (TCP, UDP or SCTP).
The software engineer writing the client and server in the application
layer only has to know about the function calls exposing the functionality
in the transport layer. This collection of functions is called the
socket API. A socket is an endpoint for networked communication; multiple
processes can open sockets via the API and data written into one socket
can be read from another. The socket API consist of functions which
can open and close the socket, send data over it and set the behavior
of a socket using 'socket options'. An example of such behavior is
the enabling or disabling of data buffering before sending to reduce
the number of packets to sent over the network and therefore improve
efficiency (Nagle's algorithm). 

\begin{figure}[h]
\caption{TCP and SCTP socket API\label{fig:TCP-and-SCTP}}

\noindent\begin{minipage}[t]{1\columnwidth}%
\begin{center}
\includegraphics[scale=0.4]{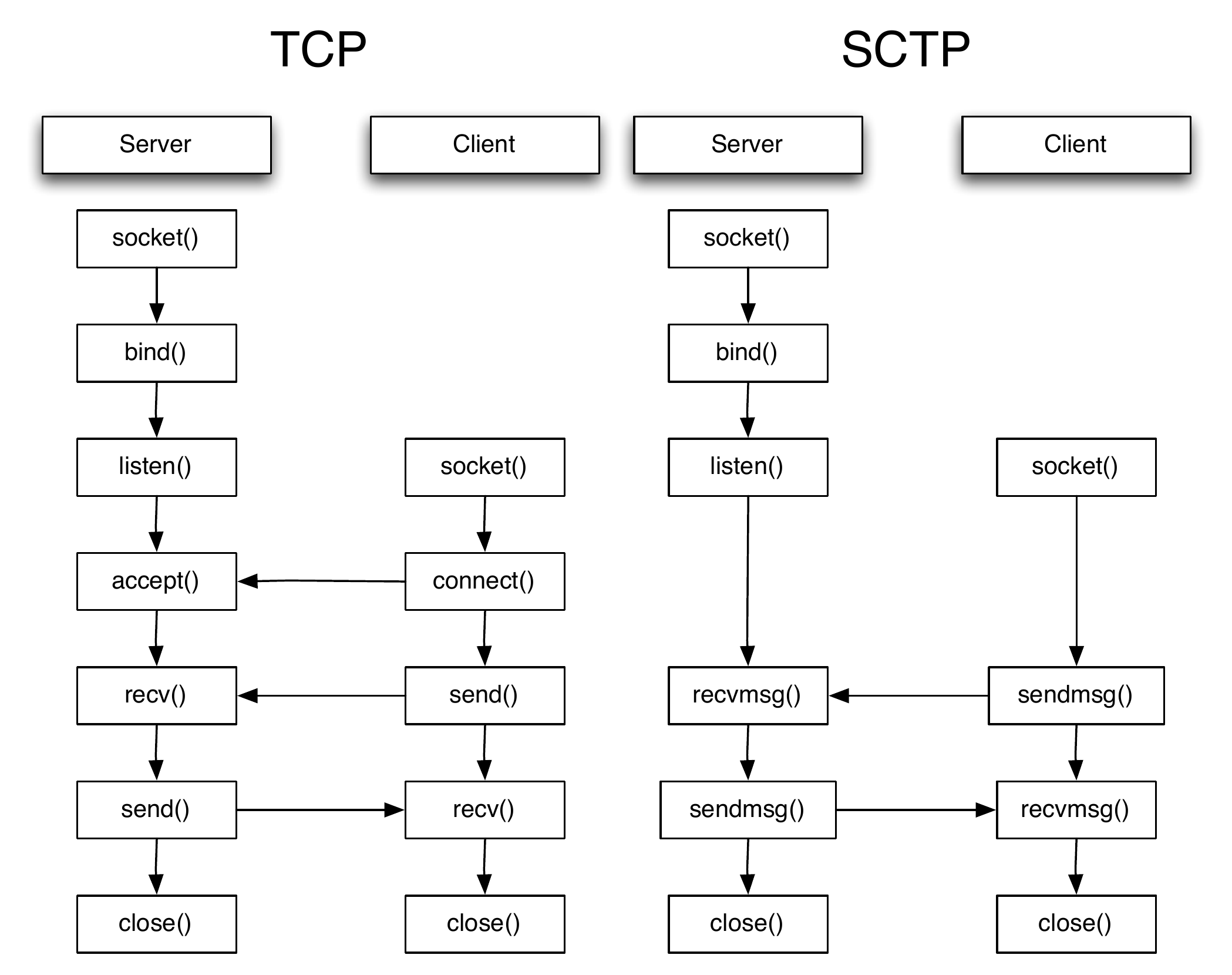}
\par\end{center}%
\end{minipage}
\end{figure}

\subsubsection{SCTP Socket API}

Figure \ref{fig:TCP-and-SCTP} is an overview of both the TCP and
SCTP socket APIs and gives the order in which the system calls would
be used in a simple client server application sending and receiving
messages. The server first creates a socket, this returns a file descriptor
which can be used by the bind function to to make a request to assign
a address to it. After this the server can start listening to incoming
connections with the listen() function. After this point TCP is different
from SCTP. The TCP client actively connects to the peer and the server
accepts the connection with the accept() and connect() functions.
With SCTP the connection set up hand handshake happens implicitly
when sending and receiving a message. At this point the server and
client can exchange messages and finally the connection terminates
with the close() function.

\begin{figure}[h]
\caption{Ancillary data embedded in sendmsg() data structure\label{fig:Ancillary-data-embedded}}

\noindent\begin{minipage}[t]{1\columnwidth}%
\begin{center}
\includegraphics[scale=0.35]{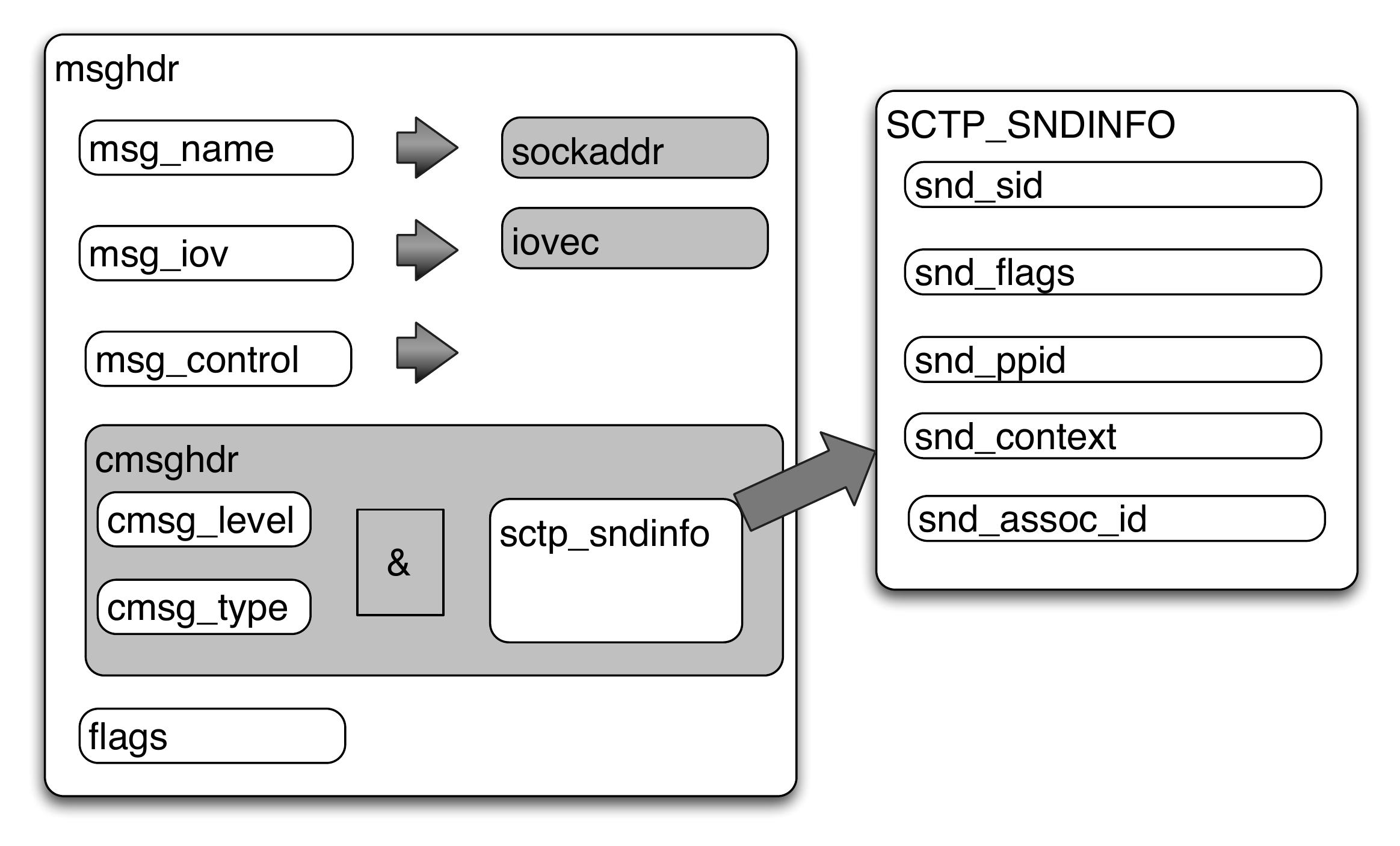}
\par\end{center}%
\end{minipage}
\end{figure}

\subsubsection{Socket I/O functions and ancillary data}

The sendmsg() and recvmsg() functions (see appendix \ref{subsec:Socket-IO-function}
for definition) are the most general of all socket I/O functions \cite[p. 390]{stevens_unix_2003}
and can be used in combination with all protocols in the transport
layer. Both functions have room for message data and ancillary data
(see appendix \ref{subsec:Message-header-structure}). SCTP adds extra
meta data which contains information about the message being sent
and its connection. Table \ref{tab:Ancillary-data-mappings} gives
an overview of the ancillary data mapped to the return variables. 

\begin{table}
\caption{Ancillary data mappings\label{tab:Ancillary-data-mappings}}
\centering{}%
\begin{tabular}{|l|l|}
\hline 
sid & stream identifier\tabularnewline
\hline 
\hline 
ssn & stream sequence number\tabularnewline
\hline 
\hline 
ppid & identifier set by peer\tabularnewline
\hline 
aid & association id\tabularnewline
\hline 
\end{tabular}
\end{table}

\subsection{Go language overview}

In this section I will give a short overview of the main features
and characteristics of the Go language and provide more detail where
relevant to the project.

\subsubsection{Data}

Go has a syntax broadly similar to the C and Java languages. Variables
are declared either by variable name followed by data type, or by
name only with data type inferred (known as 'type inference').'  With
the initialization operator := variables can be declared and initialized
in a single statement. If not initialized explicitly, the variable
is set to a default value. Here are some examples:

\begin{center}
\begin{minipage}[t]{0.8\columnwidth}%
\begin{lstlisting}[basicstyle={\scriptsize\sffamily},breaklines=true,captionpos=b,frame=tb,language=Golang,showstringspaces=false,tabsize=4]
var number int
var first_name, last_name string
var greeting = hello("olivier")
fractal := make([]uint64, 10)
\end{lstlisting}
\end{minipage}
\par\end{center}

Go has two ways to create data structures: i) with 'new', where the
data structure is initialized to zero; ii) with 'make', where the
data structure is initialized to a specified value.

\begin{center}
\begin{minipage}[t]{0.8\columnwidth}%
\begin{center}
\begin{lstlisting}[basicstyle={\scriptsize\sffamily},breaklines=true,tabsize=4,captionpos=b,frame=tb,language=Golang]
var p *[]int = new([]int)
var v  []int = make([]int, 100)
\end{lstlisting}
\par\end{center}%
\end{minipage}
\par\end{center}

An asterisk indicates that the variable is a pointer.

\subsubsection{Functions}

Unlike most commonly-used languages Go functions and methods can return
multiple values. The following function expects a string and integer
as input parameters and returns a string and an error. If the count
integer is zero the function will return a nil and an error, otherwise
it returns the greeting and nil as error.

\begin{center}
\begin{minipage}[t]{0.8\columnwidth}%
\begin{lstlisting}[basicstyle={\scriptsize\sffamily},breaklines=true,captionpos=b,frame=tb,language=Golang,tabsize=4]
func hello(name string, count int) (greeting string, err error) {
	if count = 0 {
		return nil, errors.New("Cannot say hello zero times")
	}
	greeting = "Hello" + name, nil
	return
}
\end{lstlisting}
\end{minipage}
\par\end{center}

This hello function can be be called in the following manner:

\begin{center}
\begin{minipage}[t]{0.8\columnwidth}%
\begin{lstlisting}[basicstyle={\scriptsize\sffamily},breaklines=true,captionpos=b,frame=tb,language=Golang,tabsize=4]
greeting, err := hello("paard", 0)
if err != nil {
	println("error!")
} else {
	println(greeting)
}
\end{lstlisting}
\end{minipage}
\par\end{center}

Or if we are not interested in one of the return values, a blank identifier
\_ can be used to discard the value:

\begin{center}
\begin{minipage}[t]{0.8\columnwidth}%
\begin{center}
\begin{lstlisting}[basicstyle={\scriptsize\sffamily},breaklines=true,captionpos=b,frame=tb,language=Golang,tabsize=4]
greeting, _ := hello("paard", 1)
\end{lstlisting}
\par\end{center}%
\end{minipage}
\par\end{center}

\subsubsection{Object and methods}

In Go data structures (objects) are defined as follows:

\begin{center}
\begin{minipage}[t]{0.8\columnwidth}%
\begin{lstlisting}[basicstyle={\scriptsize\sffamily},breaklines=true,captionpos=b,frame=tb,language=Golang,tabsize=4]
type Person struct {
	name string
	age int
}
\end{lstlisting}
\end{minipage}
\par\end{center}

Functionality can be added to an object by defining methods associated
with it. A difference between Go and other object-oriented languages
like Java is that in Go this functionality is defined outside of the
object, for example:

\begin{center}
\begin{minipage}[t]{0.8\columnwidth}%
\begin{lstlisting}[basicstyle={\scriptsize\sffamily},captionpos=b,frame=tb,language=Golang,tabsize=4]
func (p Person) SayHello(name String) {
	return "Hello " + name ", my name is " + p.name
}
\end{lstlisting}
\end{minipage}
\par\end{center}

In the above example the SayHello method has a string as input and
is associated with a Person object p. 

Methods can be associated with any type, not just objects. In the
following example the Add() method is associated with an Array of
strings:

\begin{center}
\begin{minipage}[t]{0.8\columnwidth}%
\begin{lstlisting}[basicstyle={\scriptsize\sffamily},breaklines=true,captionpos=b,frame=tb,language=Golang,tabsize=4]
type NameArray []string
func (na NameArray) Add(name string) []string {
	...
}
\end{lstlisting}
\end{minipage}
\par\end{center}

It's worth noting that Go has objects but not inheritance. 

\subsubsection{Interfaces }

As with other object oriented languages, you can specify behavior
of an object by using interfaces and also implement multiple interfaces
per object. The interfaces does not have to be explicitly named: as soon
as the type implements the methods of the interface the compiler knows
the relation between the two. In the following example, in function
main, a cat object gets created, it talks and then gets cast (type
changed) to an animal object:%
\begin{minipage}[t]{0.8\columnwidth}%
\begin{lstlisting}[basicstyle={\scriptsize\sffamily},breaklines=true,captionpos=b,frame=tb,language=Golang,tabsize=4]
type animal interface {   
	Talk() 
}

type Cat

func (c Cat) Talk() {   
	fmt.Println("Meow") 
}

func main() {   
	var c Cat   
	c.Talk()   
	a := animal(c)   // Cast from cat to animal
	a.Talk() 
}
\end{lstlisting}
\end{minipage}

\subsubsection{Pointers}

Although Go has pointers (references to memory locations), they are
implemented differently than in a language like C. Pointer arithmetic
is not possible so Go pointers can only be used to pass values by
reference. At a lower level Go pointers can be converted to C pointers.

\subsubsection{Concurrency}

Go has a basic primitive for concurrency called a goroutine. The name
is not only a play on a programming component called coroutine but
is also its implementation of it in the language. Coroutines are methods
which call on each other, data comes in, gets processed and gets passed
to a the next coroutine. One of the advantages of coroutines is they
are generally easier to create and understand {[}Knuth V1 p193{]}.
But the main advantage is that it lends itself very well for distributed
computing where data gets passed around from one processor to another,
maybe even on a different computer. 

In Go every function can become a goroutine by simple putting go in
front of it. A gorouting can than communicate its input and output
via channels. This concept is called Communicating sequential processes
(CSP) and is surprisingly versatile in its use. \cite{hoare_communicating_1978}
Here an example:

\begin{center}
\begin{minipage}[t]{0.8\columnwidth}%
\begin{lstlisting}[basicstyle={\scriptsize\sffamily},captionpos=b,frame=tb,language=Golang,numbers=left,tabsize=4]
package main

func receiveChan(ci chan int) {
  for {
    i := <-ci
    println(i)
  }  
}

func main() {
  ci := make(chan int)
  go receiveChan(ci)

  for i := 0; i < 10; i++ {
    ci <- i
  }  
}
\end{lstlisting}
\end{minipage}
\par\end{center}

The receiveChan() function has as input a channel of integers. On
line 4 an endless for loop starts where line 5 waits for an integer
to come in on the channel. The main functions first creates a channel
of integers. Line 12 starts the function receiveChan as a Go routine
in the background. This is followed by a loop sending 10 integers
over the channel to the receiveChan function.

\subsubsection{Much more}

There is more to the language like garbage collection, first class
functions, arrays, slices, however the explanation of this falls outside
the scope of this paper. More information can be found on the Go website
\footnote{\url{http://golang.org/doc/effective_go.html}}.

\section{Go networking}

The following section contains the findings of my research on how
Go combines the system level TCP/IP socket API into an easy-to-use
network library and how that works internally. Much of the TCP functionality
is very similar to the SCTP stack and this will serve as an example
of how to implement SCTP in Go.

\subsection{The network library}

Go provides a portable interface for network I/O. Basic interaction
is provided by the Dial, Listen, ListenPacket and Accept functions
which return implementations of the Conn, PacketConn and Listener
interfaces. The Listen function is for byte stream data connections
and the ListenPacket is for data transfer which includes messages
boundaries, like UDP or Unix domain sockets. These functions simplify
the access to the socket API, but if needed the application developer
can still access the socket API in more detail. 

Here is an example of a typical TCP client and server application.
The client sends a single message and the server waits to receive
a message, prints it out after receiving and starts listening again.
First a simple client:

\begin{center}
\begin{minipage}[t]{0.9\columnwidth}%
\begin{lstlisting}[basicstyle={\scriptsize\sffamily},breaklines=true,captionpos=b,frame=tb,language=Golang,numbers=left,numberstyle={\scriptsize},tabsize=4]
package main
import "net"

func main() {
	conn, err := net.Dial("tcp", "localhost:1234")
	if err != nil {
		return
	}
	defer conn.Close()
	conn.Write([]byte("Hello world!"))
}
\end{lstlisting}
\end{minipage}
\par\end{center}

The function main() is the entry point of the program. The first step
the client performs is 'dialing' a server with the TCP protocol (line
5). The Dial() function returns a connection which is an implementation
of the Conn interface. After checking for errors (6-8) the defer keyword
before the connection close command indicates the connection can be
finished as soon as it is no longer needed. In this case it happens
immediately after the write so it does not make much sense to defer
it, but in larger programs with multiple exit points you only need
a single (deferred) close statement, which makes it easier to understand
and maintain the program.

Next the server:

\begin{center}
\begin{minipage}[t]{0.9\columnwidth}%
\begin{lstlisting}[basicstyle={\scriptsize\sffamily},breaklines=true,captionpos=b,frame=tb,language=Golang,numbers=left,numberstyle={\scriptsize},tabsize=4]
package main 
import "net"

func main() {
	listen, err := net.Listen("tcp", "localhost:1234")
	if err != nil {
		return
	}
	buffer := make([]byte, 1024)
	for {
		conn, err := listen.Accept()
		if err != nil {
			continue
		}
		conn.Read(buffer)
		println(string(buffer))
	}
} 
\end{lstlisting}
\end{minipage}
\par\end{center}

The server gets created with the Listen() method and returns an object
which implements the Listener interface. On line 9 a byte array buffer
gets created for the received data. The following for loop (10 - 17)
will continuously wait to accept an incoming connection (11), check
for errors after connect, read from it (15) and convert the byte array
to a string before printing it (16). 

\subsection{Under the hood\label{subsec:Under-the-hood}}

Go's network communication library uses the same socket API as any
C program. In this section I will examine what happens when a network
connection is set up and which socket API calls are made at what point.
The example uses TCP. To illustrate how Go accesses the socket API
I will take the TCP client described in the previous section as an
example and show how it interacts with the kernel by opening a connection. 

\subsubsection*{Socket and Connect}

For a TCP client to create a connection and send some data the following
system calls need to be made in this order:
\begin{enumerate}
\item resolve IP address
\item socket()
\item setsockopt() (optional)
\item connect()
\end{enumerate}
In Go all this is wrapped in the net.Dial() call. 

Figure \ref{fig:TCP-Client-setting} shows a sequence diagram of method
calls after a client calls Dial(). First the hostname:port gets parsed
and resolved to an IP address (1.1)\footnote{There are more method calls behind this but they are not relevant
for this example}. Net.dialAddr() (1.3) determines the transport protocol type and
calls the relevant method, net.DialTCP() in this case (1.3.1). Next
net.internetSocket() gets called which internally calls socket() in
the syscall package. Syscall.socket() is an auto-generated method
based on C header files which describe the socket API.

Every network protocol in Go has its own connection type. As you can
see in figure \ref{fig:TCP-Client-setting} the generic Dial() method
eventually reaches the specific DialTCP() method which returns a TCP-
specific connection type. This type gets cast to the more generic
Conn type which can be used in the client application. If TCP-specific
functionality is needed the Conn type can be recast to a TCPConn which
then makes it possible to access TCP-specific functionality. 

\subsection{Auto-generation}

Native Go functions to access kernel system calls and data structures
can be auto-generated via scripts. In FreeBSD, if the description
of the system call is present in the syscalls.master file \footnote{System call name/number master file: http://fxr.watson.org/fxr/source/kern/syscalls.master},
the function will be available through the syscall package. Most system
calls which are used in Go are wrapped in methods to make them fit
better the language. The structures which are passed to the system
calls are created in the same way. The source code directory of the
syscall package contains a file with a link pointing to a header file
which describes the structures. 

\subsection{Go non-blocking networking }

All networking in Go is non blocking which is not the default in the
TCP/IP model. As soon as a application tries to retrieve data from
a socket, e.g. via readmsg(), it will not block until some data comes
in, instead it will immediately move on in the program. Reading data
from the socket is normally done in an endless loop. To make Go actually
wait for sending data back to the application Go implements the reactor
design pattern. The implementation of this software pattern makes
use of a mechanism in the kernel where you can ask the kernel to send
a signal back to the application if certain events occur whilst you
keep doing other things in the program. For example data being written
to the socket descriptor. On BSD the system call kqeueu() and kevent()
are used to register and queue events. The functionality of this is
handled by Go's runtime. 

\section{Design}

In the following section I will describe how the different Go network
methods will map to the SCTP socket API. 

\begin{figure}[H]
\caption{Go SCTP network API\label{fig:Go-SCTP-network}}

\noindent\begin{minipage}[t]{1\columnwidth}%
\begin{center}
\includegraphics[scale=0.45]{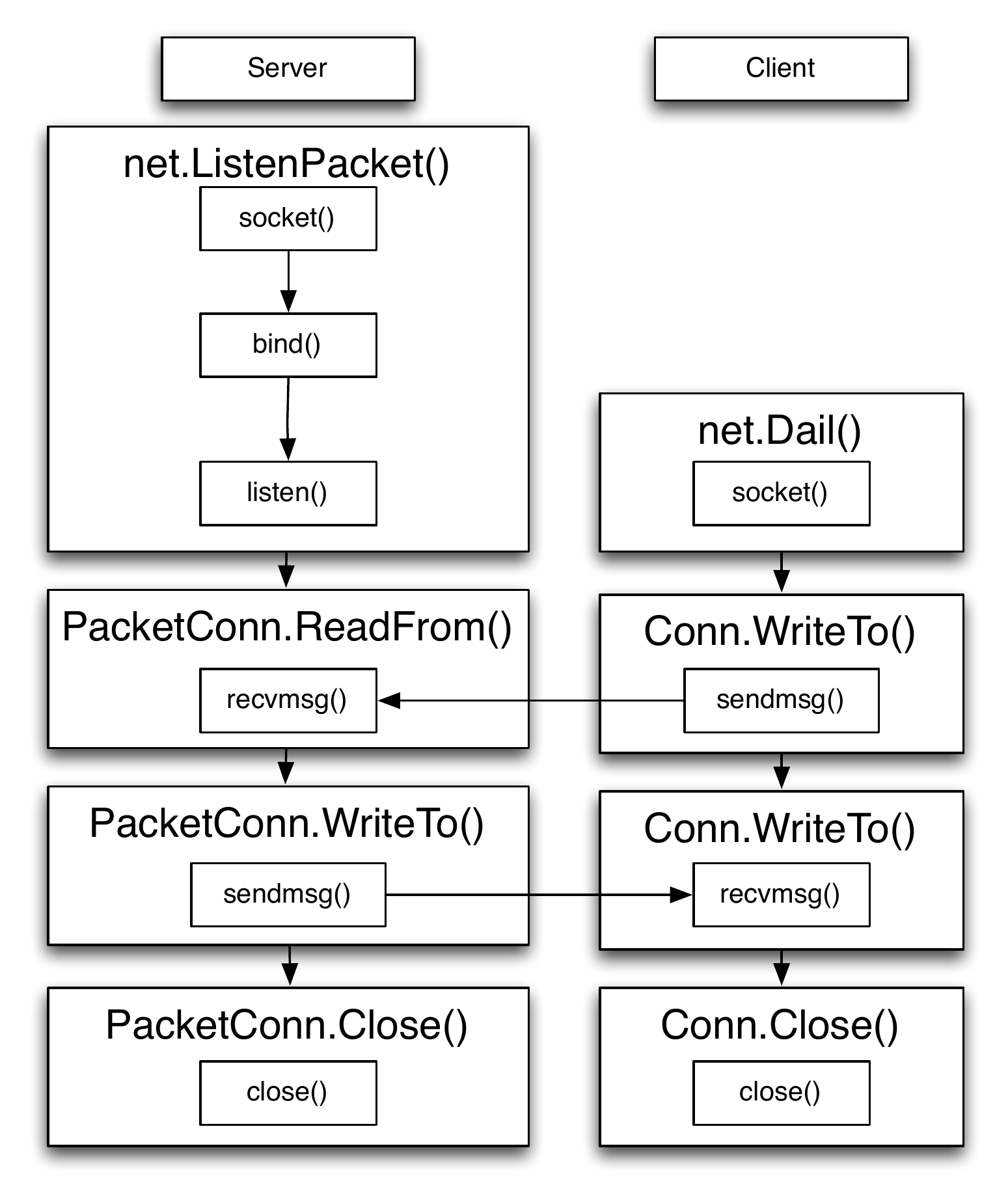}
\par\end{center}%
\end{minipage}
\end{figure}

\subsection{Mapping APIs \label{subsec:Mapping-APIs}}

The SCTP socket API will follow closely the wrapping of the TCP socket
API in the Go network library as described in section \ref{subsec:Under-the-hood}.
Figure \ref{fig:Go-SCTP-network} show the mapping of the socket API
to the different functions and methods. At the server side the socket(),
bind() and listen() functions are bundled into the ListenPacket()
function which resides in the network package (net). The ListenPacket()
function returns an implementation of the PacketConn interface (See
appendix \ref{subsec:PacketConn-interface}). The client wraps the
socket() function into the Dail() function. This functions returns
a Conn (connection) interface which can used for writing messages
(Conn.WriteTo()) and these user messages can be received vie the ReadFrom()
method on the PacketConn interface.

\subsection{SCTP specific }

\subsubsection{Receive SCTP specific information }

To access SCTP specific functionality, such as which stream the message
has been sent on, or the association id, net.ListenSCTP() can be used.
This method returns a SCTP specific type (SCTPConn) which has the
method ReadFromSCTP() and WriteToSCTP() added to it. These methods
return and set the information contained by the SCTP receive information
structure, added as ancillary data when the system call recvmsg()
returns. 

\subsubsection{Send SCTP specific information}

To be able to set SCTP specific send information such as stream id
or association id via the SCTP Send Information Structure, the several
methods on the SCTPConn object can be used (See table \ref{tab:Initialization-parameters}):

\begin{table}[H]
\caption{Initialization parameters\label{tab:Initialization-parameters}}
\centering{}%
\begin{tabular}{|l|l|}
\hline 
{\scriptsize{}Number of output streams} & {\scriptsize{}({*}SCTPConn) InitNumStreams(n int) error}\tabularnewline
\hline 
{\scriptsize{}Max number of input streams } & {\scriptsize{}({*}SCTPConn) InitMaxInStream(n int) error}\tabularnewline
\hline 
{\scriptsize{}Max number of attempts to connect } & {\scriptsize{}({*}SCTPConn) InitMaxAttempts(n int) error}\tabularnewline
\hline 
{\scriptsize{}Timeout} & {\scriptsize{}({*}SCTPConn) InitMaxInitTimeout(n int) error}\tabularnewline
\hline 
\end{tabular}
\end{table}

A typical server which has access to SCTP specific functionality would
look like this:

\begin{center}
\noindent\begin{minipage}[t]{1\columnwidth}%
\begin{lstlisting}[basicstyle={\scriptsize\sffamily},breaklines=true,captionpos=b,frame=tb,language=Golang,tabsize=4]
package main 
import (   
    "net"
    "strconv" 
)

func main() { 
   addr, _ := net.ResolveSCTPAddr("sctp", "localhost:4242") 
   conn, _ := net.ListenSCTP("sctp", addr)   
   defer conn.Close()   
   for {     
      message := make([]byte, 1024)
       _, _, stream, _ := conn.ReadFromSCTP(message)     
       println("stream " + strconv.Itoa(int(stream)) + ": " + string(message))   
    }   
}
\end{lstlisting}
\end{minipage}
\par\end{center}

In this program ListenSCTP returns a SCTP connection type. This type
implements Conn and PacketConn interface and has the ReadFromSCTP
method added to it. The println() functions prints the stream id and
the user message.

\section{Implementation}

In this section I will describe how the SCTP network functionality
can fit into the existing Go network framework. The main goal of the
SCTP application programming interface (API) design is to combine
lower-level system calls in an easy-to-use framework. This framework
hides the underlying complexity of the socket API but if needed gives
access to all the functionality provided by the protocol. To make
sure SCTP fits in the Go design philosophy, less is more, I will make
use as much as possible of the existing components and interfaces
in the Go network package. In the following section I'll e

\subsection{Server}

For a server to be able to set up a SCTP association it needs to create
a socket, bind to it, optionally set some socket options and start
listening to it. A typical server will access the socket API in the
following sequence:
\begin{enumerate}
\item socket()
\item bind()
\item listen()
\item recvmsg()
\end{enumerate}
The socket(), bind() and listen() functions will be wrapped into a
Listen() method which returns a connection type. There are three variations:
net.Listen(), net.ListenPacket() and net.ListenSCTP(). The Go network
library provides the net.ListenPacket() method for packet-oriented
networking like UDP, and net.Listen() for stream-oriented networking
like TCP. SCTP which is packet-oriented, can also make use of the
net.ListenPacket() method. ListenPacket() returns an implementation
of the PacketConn interface which can be used to read and write messages
(recvmsg()). A simple SCTP echo server using the PacketConn interface
might look like this:

\begin{center}
\begin{minipage}[t]{0.9\columnwidth}%
\begin{lstlisting}[basicstyle={\footnotesize\sffamily},breaklines=true,captionpos=b,frame=tb,language=Golang,tabsize=4]
package main 
import "net"

func main() {   
   conn, _ := net.ListenPacket("sctp", "localhost:4242")   
   defer conn.Close()   
   message := make([]byte, 1024)    
   conn.ReadFrom(message)   
   print(string(message)) 
}
\end{lstlisting}
\end{minipage}
\par\end{center}

After the the main entry point a connection object is created via
the ListenPacket() method. The parameters of this method indicate
that the connection should use the SCTP protocol and listen on localhost
port 4242. The next line defers the closing of the connection when
it is not needed anymore. After creating a byte array buffer to store
incoming data a message is read from the connection. The ReadFrom()
method will block until a message is received. Finally the message
is printed and the program ends. 

\subsubsection*{Receive SCTP-specific information}

To access SCTP-specific functionality, such as which stream the message
has been sent on, or the association id, net.ListenSCTP() can be used.
This method returns a SCTP-specific type (SCTPConn) which has the
method ReadFromSCTP() added to it:

\begin{center}
\begin{minipage}[t]{0.9\columnwidth}%
\begin{lstlisting}[basicstyle={\footnotesize\sffamily},breaklines=true,captionpos=b,frame=tb,language=Golang,tabsize=4]
(*SCTPConn).ReadFromSCTP(message *string)(sid int, ssn int, ppid int, aid int, addr SCTPAddr, err error)
\end{lstlisting}
\end{minipage}
\par\end{center}

The ReadFromSCTP() method returns the information contained by the
SCTP receive information structure, added as ancillary data when the
system call recvmsg() returns. A typical server which has access to
SCTP-specific functionality would look like this:

\begin{center}
\begin{minipage}[t]{0.9\columnwidth}%
\begin{lstlisting}[basicstyle={\scriptsize\sffamily},breaklines=true,captionpos=b,frame=tb,language=Golang,tabsize=4]
package main 
import (   
    "net"
    "strconv" 
)

func main() { 
    addr, _ := net.ResolveSCTPAddr("sctp", "localhost:4242") 
    conn, _ := net.ListenSCTP("sctp", addr)   
    defer conn.Close()   
    for {     
       message := make([]byte, 1024)
       _, _, stream, _ := conn.ReadFromSCTP(message)     
       println("stream " + strconv.Itoa(int(stream)) + ": " + string(message))   
    }   
}
\end{lstlisting}
\end{minipage}
\par\end{center}

In this program ListenSCTP() returns a SCTP connection type. This
type implements the Conn and PacketConn interfaces and adds the ReadFromSCTP()
method. 

\subsection{Client}

In Go a client connection sets itself up with a call to the Dial()
method. The Dial() method returns the generic Conn interface. Every
network protocol in Go has its own Dial() method which returns a protocol-specific
connection type. In the case of SCTP this is the PacketConn type which
has underneath it a specific SCTP connection type (SCTPConn). A simple
SCTP client sending a single message would look like this:

\begin{center}
\begin{minipage}[t]{0.9\columnwidth}%
\begin{lstlisting}[basicstyle={\scriptsize\sffamily},breaklines=true,captionpos=b,frame=tb,language=Golang,tabsize=4]
package main
import "net"

func main() {   
    addr, _ := net.ResolveSCTPAddr("sctp", "localhost:4242")    
    conn, _ := net.DialSCTP("sctp", nil, addr)
    defer conn.Close()
    message := []byte("paard")
    conn.WriteTo(message, addr)
}
\end{lstlisting}
\end{minipage}
\par\end{center}

The DialSCTP() method creates the socket and sets default socket options.
Sending the message via WriteTo() will implicitly set up the connection. 

\subsubsection*{Send SCTP-specific information}

To be able to set SCTP-specific send information such as stream id
or association id via the SCTP Send Information Structure, the WriteToSCTP()
method can be used:

\begin{center}
\begin{minipage}[t]{0.9\columnwidth}%
\begin{lstlisting}[basicstyle={\scriptsize\sffamily},breaklines=true,captionpos=b,frame=tb,language=Golang,tabsize=4]
(*SCTPConn).WriteToSCTP(message *string, addr SCTPAddr, sid int, ssn int, ppid int, aid int, err error)
\end{lstlisting}
\end{minipage}
\par\end{center}

\subsubsection*{Creating and binding of the socket }

The sequence diagram in figure \ref{fig:Creating-and-bind} gives
an overview of how a socket is created and bind to. At point 1.1.3
in this figure net.internetSocket() returns a socket descriptor which
is used to create the actual SCTP connection type. At this point the
SCTP initialization structure is set to its default values together
with the NODELAY socket option. The 'no delay' socket option disables
buffering of bytes sent over the connection. This is also the default
setting for the TCP implementation in Go.

\section{Analysis}

\subsection{Performance}
For performance testing a client-server application was
designed, with the client sending messages and the server
receiving them. Both a C and a Go version of this application
were created and compared and run against each other. The
data throughput of an SCTP client-server application written in
Go is approximately twice as high as the same program written
in C. Most of the delay happens in the Go client. Unlike the C
implementation, the Go application needs to go through some
additional method calls before it reaches the system call which
sends the data. Go is also garbage-collected, which causes an
extra delay because several data structures are redeclared each
time a message is sent. Another factor to consider is that the
implementation of SCTP in Go is against version 1.0.2 of
Go. This version does not have a compiler which is good at
optimizing. Newer versions of Go address this issue.

\subsection{Fitting like a glove}

Since Go already provided methods and interfaces for message based
data transmission which could be reused and because of it similarity
with the TCP socket API making SCTP available in the Go network library
was a relatively straightforward task. I could reuse of the ancillary
data structure from the Unix socket API and I only had to add the
SCTP specific ancillary data to the structure. It was easy to follow
the design philosophy 'less is exponentially more': the SCTP socket
API which is very similar to TCP could be wrapped in an easier to
use API just like it is done with TCP. This resulted in a Go SCTP
API which can be used in the most simple way hiding all complexity
of the protocol or if needed it is possible to dive deeper and make
use of more specific SCTP functionality.

\subsection{Issues during implementation}

During the implementation there where two major changes which caused
an unexpected setback. The first as mentioned before was the changing
SCTP socket API which made large parts of the implementation I already
had in place obsolete. This forced me to rewrite the majority of the
implementation. The second issue which proved to be a challenge was
the first release (1.0) of Go. Until this release I only sporadicly
synced my version of Go with the latest changes of the main development
tree of the canonical Go source code repository. Building up to the
new version the Go development team did a considerable amount of work.
So with the 1.0 release a large amount of changes needed to be incorporated
into my own branch. Since there where lots of changes in the internals
of Go I had to deal with a large amount of merge conflicts in certain
areas, specifically around the implementation of the generic Dial
and Listen interfaces. Most of the work in this area had to be redone. 

\subsection{Extensions}

There are many extensions to SCTP described in multiple RFCs. A complete
SCTP implementation should include \cite{SCTP_what_how}: 
\begin{enumerate}
\item RFC4960 (basic SCTP)
\item RFC3758 (partial reliability)
\item RFC4895 (authentication)
\item RFC5061 (dynamic addresses) 
\end{enumerate}
The last three in this list have not be included in this implementation.

\section{Conclusion}

Because of its similarity with existing protocols available in the
Go networking library SCTP fits easily into the existing Go network
library. The biggest challenge implementing this was the ongoing work
on the SCTP specification and Go itself which made the implementation
of this project a moving target. Only recently (mid way 2012) the
APIs of Go and SCTP have been stabilized. It should be noted however
that there are many extensions to SCTP described in multiple RFCs.
The work in this paper only looked at a bare minimum which is needed
to make SCTP work in Go. 

\subsection{Future work}

\subsubsection*{User land implementation of SCTP in Go}

Not all operating systems support the SCTP protocol natively. It is
however possible to have SCTP running on top of the UDP protocol,
outside the kernel (user land). To make this work a user land implementation
of SCTP on top of UDP needs to be added to Go. Once this is done SCTP
could become available on all different platforms supported by the
language.

\bibliographystyle{plain}
\bibliography{SCTP}

\appendix

\section{Go network types and interfaces\label{sec:Network-types-and}}

\subsection{Conn interface\label{subsec:Conn-interface}}
\begin{center}
\begin{minipage}[t]{0.8\columnwidth}%
\begin{center}
\begin{lstlisting}[basicstyle={\scriptsize\sffamily},language=Golang,tabsize=4]
type Conn interface {
    Read(b []byte) (n int, err error)

    Write(b []byte) (n int, err error)

    Close() error

    LocalAddr() Addr

    RemoteAddr() Addr

    SetDeadline(t time.Time) error

    SetReadDeadline(t time.Time) error

    SetWriteDeadline(t time.Time) error
}
\end{lstlisting}
\par\end{center}%
\end{minipage}
\par\end{center}

\subsection{PacketConn interface\label{subsec:PacketConn-interface}}
\begin{center}
\begin{minipage}[t]{0.8\columnwidth}%
\begin{lstlisting}[basicstyle={\scriptsize\sffamily},language=Golang,tabsize=4]
type PacketConn interface {

    ReadFrom(b []byte) (n int, addr Addr, err error)

    WriteTo(b []byte, addr Addr) (n int, err error)

    Close() error

    LocalAddr() Addr

    SetDeadline(t time.Time) error

    SetReadDeadline(t time.Time) error

    SetWriteDeadline(t time.Time) error
}
\end{lstlisting}
\end{minipage}
\par\end{center}

\subsection{Listener interface}
\begin{center}
\begin{minipage}[t]{0.8\columnwidth}%
\begin{lstlisting}[basicstyle={\scriptsize\sffamily},language=Golang,tabsize=4]
type Listener interface {
    Accept() (c Conn, err error)

    Close() error

    Addr() Addr
}
\end{lstlisting}
\end{minipage}
\par\end{center}

\begin{figure*}[tp]
\caption{TCP Client setting up connection\label{fig:TCP-Client-setting}}
\raggedright{}%
\noindent\begin{minipage}[t]{1\columnwidth}%
\begin{center}
\includegraphics[clip,scale=0.9]{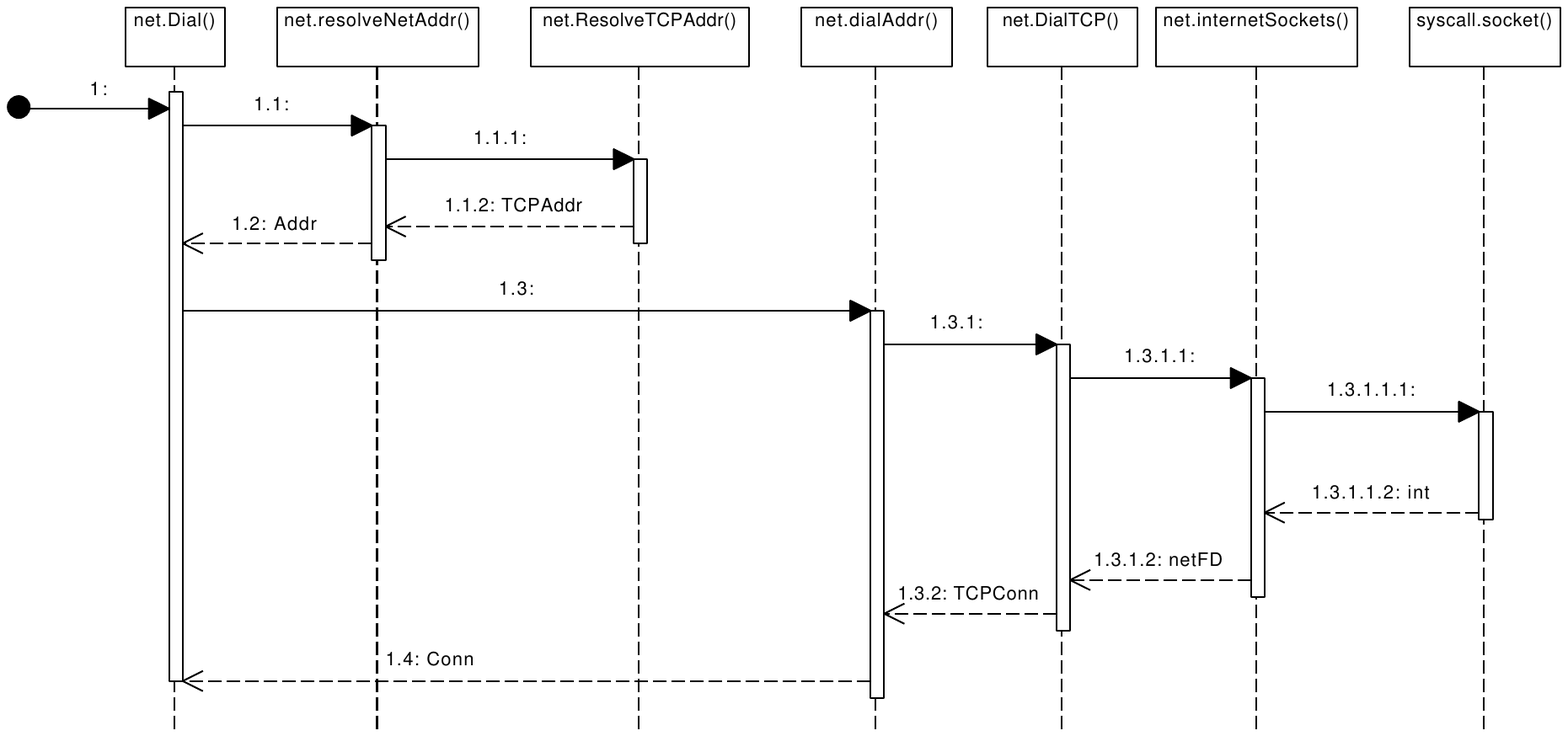}
\par\end{center}%
\end{minipage}
\end{figure*}

\begin{figure*}[p]
\begin{centering}
\caption{Creating and bind\label{fig:Creating-and-bind}}
\par\end{centering}
\noindent\begin{minipage}[t]{1\columnwidth}%
\begin{center}
\includegraphics[scale=0.9]{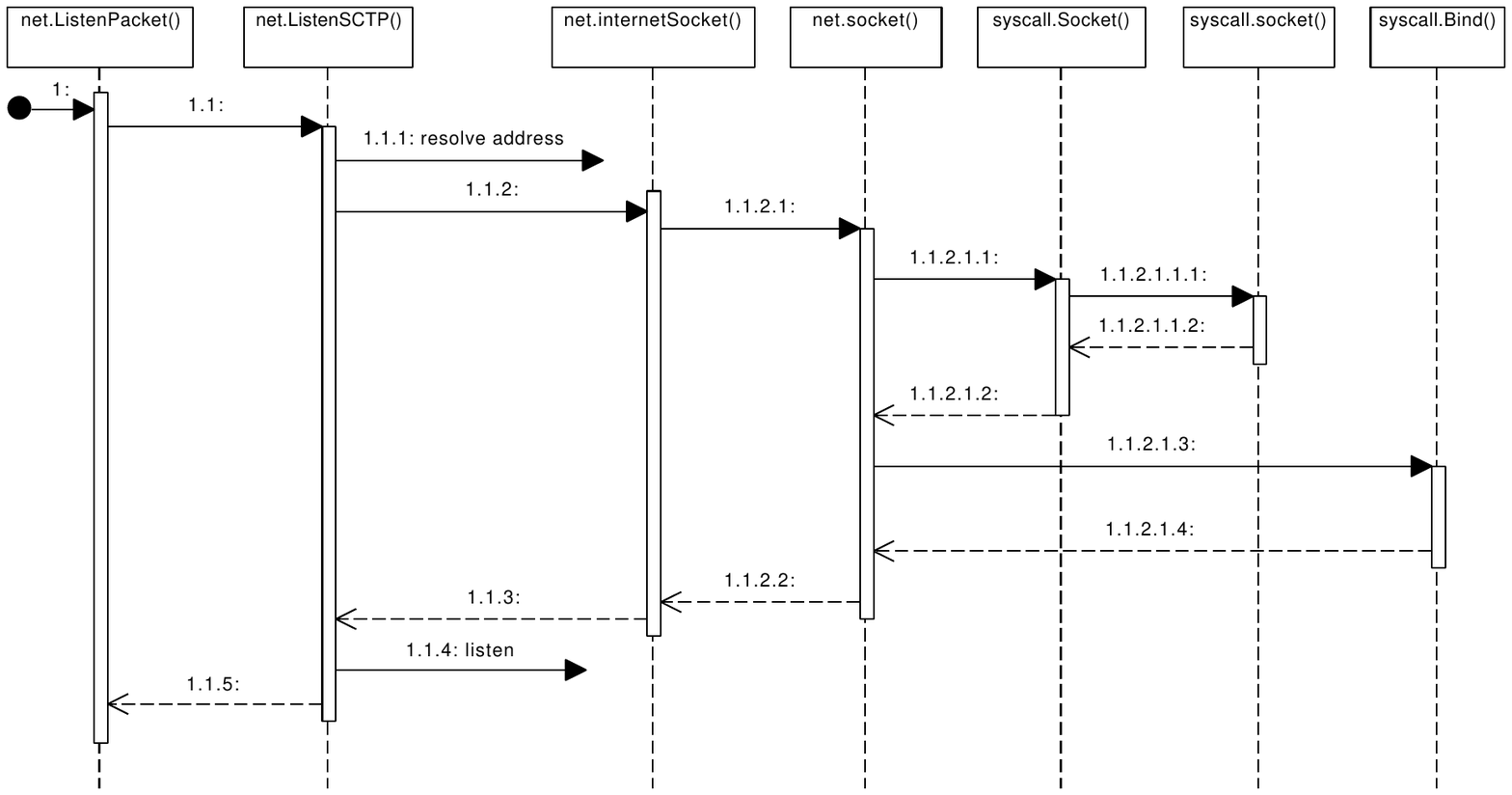}
\par\end{center}%
\end{minipage}
\end{figure*}

\section{Socket IO functions}

\subsection{Socket IO function definition\label{subsec:Socket-IO-function}}

\begin{minipage}[t]{0.8\columnwidth}%
\begin{lstlisting}[basicstyle={\scriptsize\sffamily},language=C,tabsize=4]
ssize_t sendmsg(int s, const struct msghdr *msg, int flags)
\end{lstlisting}
\end{minipage}

\begin{minipage}[t]{0.8\columnwidth}%
\begin{lstlisting}[basicstyle={\scriptsize\sffamily},language=C,tabsize=4]
ssize_t recvmsg(int s, struct msghdr *msg, int flags)
\end{lstlisting}
\end{minipage}

\subsection{Message header structure and ancillary data \label{subsec:Message-header-structure}}

\noindent\begin{minipage}[t]{1\columnwidth}%
\begin{lstlisting}[basicstyle={\scriptsize\sffamily},language=C,tabsize=2]
struct msghdr {              
  void         *msg_name;      /* optional address */
  socklen_t     msg_namelen;   /* size of address */
  struct iovec *msg_iov;       /* scatter/gather array */
  int           msg_iovlen;    /* # elements in msg_iov */
  void         *msg_control;   /* ancillary data */
  socklen_t     msg_controllen;/* ancillary data buffer len */             
  int           msg_flags;     /* flags on message */      
};
\end{lstlisting}
\end{minipage}

The msg\_control argument, which has length msg\_controllen, points
to a buffer for other protocol control related messages or other miscellaneous
ancillary data. The messages are of the form:%
\noindent\begin{minipage}[t]{1\columnwidth}%
\begin{lstlisting}[basicstyle={\scriptsize\sffamily},language=C,tabsize=4]
struct cmsghdr {
  socklen_t  cmsg_len;    /* data byte count, including hdr */              
  int        cmsg_level;  /* originating protocol */              
  int        cmsg_type;   /* protocol-specific type */      
 /* followed by u_char cmsg_data[]; */
};
\end{lstlisting}
\end{minipage}
\end{document}